\documentclass{article}
\usepackage{amssymb}
\usepackage{amsfonts}
\usepackage{times}
\usepackage{color}
\usepackage{graphicx}
\usepackage{esint}
\usepackage{longtable}
\usepackage{dcolumn}

\usepackage{xparse}

\begin{document}

\title{Maximum Force in Modified Gravity Theories}
\author{John D. Barrow$^{1}$ and Naresh Dadhich$^{2}$ \\
%EndAName
$^{1}$DAMTP, Centre for Mathematical Sciences,\\
University of Cambridge, Cambridge CB3 0WA, U.K.\ \ \ \\
$^{2}$\ IUCAA, Post Bag 4, Ganeshkhind, Pune 411007, India}
\maketitle
\date{}

\begin{abstract}
We investigate the existence and nature of classical maximum force bound between two black holes with touching horizons. Besides general relativity, the maximum force bound is independent of black hole masses only in Moffat's theory, Brans Dicke theory (which is the same as Einstein's for vacuum) and the higher dimensional generalization of Einstein's theory, pure Lovelock gravity which is characterised by having single $n$th order term in Lovelock polynomial without sum over lower orders in the action. Further if the bound is to exist in higher dimensions and is entirely in terms of the velocity of light and the gravitational constant, it has uniquely to be pure Lovelock gravity. In pure Lovelock gravity, the maximum force bound exists in all $3n$-space dimensions and has the value, $c^4/4G_n$ where $G_n$ is the corresponding gravitational constant. The absence of mass dependence in the maximum force bound may have relevance for the formation of naked singularities. 

%We investigate the existence and nature of classical maximum force constraints in some gravity theories other than Einstein's. We calculate the maximum force bounds in several of these gravity theories. We are interested in mass-independent maximum force bounds between black holes with touching horizons and find that this mass-independent feature of general relativity in three dimensions is present in Moffat's gravity theory, Brans Dicke theory, and the pure Lovelock theory in which only one term in the lagrangian sum for the general Lovelock lagrangian is retained but not the sum over lower-order terms. It is the only theory that has this property in higher dimensions when the spatial dimension is equal to three times the Lovelock order contributing the single lagrangian to define the Lovelock action. We determine whether the maximum force is bigger or smaller than in general relativity. The absence of mass dependence in the maximum force relations may have relevance for the formation of naked singularities.
\end{abstract}

\section{Introduction}

It has been proposed and demonstrated in a wide range of situations, \cite%
{Gibb2002, JBGib} that in general relativity (GR) there should be a maximum
value to any physically attainable force, or tension, given by
\begin{equation}
F_{\max }=\frac{c^{4}}{4G}\,,  \label{1}
\end{equation}%
where $c$ is the velocity of light and $G$ is the Newtonian gravitational
constant. This motivates closely related conjecture that there is a maximum
power defined by

\begin{equation}
P_{\max }=cF_{\max }=\frac{c^{5}}{4G},  \label{2}
\end{equation}%
the so-called Dyson Luminosity \textit{\cite{dyson}},\textit{\ }or some
multiple of it to account for geometrical factors $O(1)$. This limits
maximum possible luminosity in gravitational waves, or indeed other forms of
radiation that an isolated system may emit, \cite{schiller}, \cite{sperhake}%
. Schiller has come to the same conclusion and proposed a stronger thesis:
that the existence of a maximum force implies general relativity, just as a
maximum velocity characterises special relativity. \footnote{This is however not quite true as we shall see in the following that pure Lovelock theory as well as Moffat's gravity theory do admit maximum force. What seems to be true though is that existence of maximum force bound in terms of only the fundamental constants, the velocity of light and the gravitational constant leads to pure Lovelock gravity which includes GR at the linear order $n=1$.} This claim is much less
clear since it requires in effect a proof of cosmic censorship. It is also
necessary to choose quite subtle energy conditions in order to avoid the
formation of sudden singularities \cite{Barrow2004, Barrow2004a, Barrow2004b}%
, where unbounded pressure forces (or its time-derivatives) will occur. The
origin of the maximum force and luminosity bounds lies in the fact that the
Planck units of these quantities do not contain Planck's constant: they are
entirely classical and exist in the presence of a cosmological constant \cite%
{JBGib}. In $N$-dimensional space, the Planck unit of force in powers of the
fundamental constants $G,c,$ and $h$ is%
\begin{equation}
F_{pl}=G^{2/(1-N)}c^{(5+N)/(N-1)}h^{(3-N)/(1-N)},  \label{force}
\end{equation}%
and we see the disappearance of Planck's constant of action, $h$, when and
only when $N=3$.  A further example, the magnetic moment to angular momentum,
has also been identified \cite{BarGib2}. This signals something fundamental
about these non-quantum natural units that is unique to three-dimensional
space \cite{JBsing}. Note that this bound on forces does not exist in
Newtonian gravity \cite{JBnewt}, where point masses can approach arbitrarily
closely and the inverse-square gravitational force can become arbitrarily
large. It is the formation of an event horizon around these mass points in
general relativity that gives rise to the maximum force: it corresponds to
the inverse square law force between two Schwarzschild black holes of the
same mass touching at their horizons. If their masses are unequal an
inequality ensures bounded by the maximum force \cite{JBsing}.

The important point to note is that though black hole horizon exists in all higher dimensions but mass independent upper bound on maximum force exists only in three space dimensions. %bound between two black holes in higher-dimensional extensions of GR is \cred{independent of the masses;i.e. mass-independent only} in three-dimensional spaces.
We will investigate whether this feature is shared in Lovelock's
generalization of GR in higher dimensions.

In this paper we will extend these analyses of the existence of a maximum
force to some other gravity theories where static spherically symmetric
solutions are available to do an analogous exact calculation of the sort
performed in GR. We identify theories that give the same
mass-independent maximum force as GR. We set the speed of light, $c,$ equal
to unity unless we specify otherwise.

\section{Gravity Theories Beyond Einstein}

In this section we will study the existence of maximum forces and compare
them with the bound derived for general relativity. In general, we will
evaluate the magnitude of the force between two
vacuum static spherically symmetric black holes that touch horizons .

\subsection{Moffat's theory}

Moffatt's extension of GR \cite{moffatt, sporea} has a black hole solution
with static spherically symmetric metric:

\begin{equation}
ds^{2}=h(r)dt^{2}-\frac{dr^{2}}{h(r)}-r^{2}d\Omega ^{2},  \label{1a}
\end{equation}

where

\begin{equation}
h(r)=1-\frac{2(1+\alpha )G_{N}M}{r}+\frac{\alpha (1+\alpha )M^{2}G_{N}^{2}}{%
r^{2}},  \label{a}
\end{equation}

where

\begin{equation}
G=G_{N}(1+\alpha ),  \label{b}
\end{equation}

with $G$ the effective gravitation constant, and $G_{\rm{N}} $ the
Newtonian constant (the speed of light is unity).  This solution approaches the Schwarzschild metric of GR as $\alpha\rightarrow 0$. Its event horizons are at

\begin{equation}
R_{g}^{\pm }=G_{N}M\left[ 1+\alpha \pm \sqrt{1+\alpha }\right] ,  \label{c}
\end{equation}%
and we take the positive sign to obtain the GR result as $\alpha \rightarrow
0$.

The force between two equal-mass black holes of mass $M$ ($(1/2)Mh'$) touching at this radius for each is
%\footnote{%The bound is stronger for unequal masses: to see this, use the inequality $(%\textcolor{red}{\sqrt{M_{1}}}-\sqrt{M_{2}})^{2}=M_{1}+M_{2}-2\sqrt{M_{1}M_{2}%
%}$\geq 0.$%\parHence, $M_{1}M_{2}\leq \frac{1}{4}(M_{1}+M_{2})^{2}$ in the formula for $%F_{N}(M_{1},M_{2})$ and we find for the maximum force $F_{N}(M_{1},M_{2})%\leq F_{N}(M_{1},M_{1})$ always.}

\begin{equation}
F_{moff}=\frac{(1+\alpha)M\bar M}{r^2}(1-\frac{\alpha\bar M}{r})
%\frac{(1+\alpha )G_{N}M^{2}}{(R_{g}^{+})^{2}}=\frac{(1+\alpha )}{%%
%G_{N}\left( 1+\alpha +\sqrt{1+\alpha }\right) ^{2}}.  \label{d}
\end{equation}%
where $\bar M = G_{N}M$. This when evaluated at the horizon, $R_{g}^{+}$, and on restoring the speed of light reads as follows:

\begin{equation}
F{moff} = \frac{c^4}{G}\frac{\beta}{(\beta+\sqrt{\beta})^3}(\beta+\sqrt{\beta}-\alpha)
\end{equation}

where $\beta=1+\alpha$. This reduces to the GR result when $\alpha
=0:$

\begin{equation}
F_{GR}=\frac{c^{4}}{4G_{N}}.  \label{e}
\end{equation}

However, when $\alpha >0$, it can be less than the GR bound. For example, at large $\alpha ,$
\begin{equation}
F_{moff}\rightarrow \frac{c^{4}}{\alpha^{3/2} G_{N}}<F_{GR}.  \label{f}
\end{equation}

For small $\alpha >0$, we have

\begin{equation}
F_{moff}\rightarrow F_{GR}\left( 1-\frac{5\alpha }{8}+O(\alpha ^{2}\right)
<F_{GR}.  \label{g}
\end{equation}%}
In general, the maximum force in eq.(\ref{e}) arises when $\alpha =0$ and $%
F_{moff}<F_{GR}$ for all $\alpha >0.$

\subsection{Brans-Dicke Theory}

There are four varieties of static spherically symmetric vacuum metrics in
Brans-Dicke (BD) gravity \cite{BD, camp, bad} and only one is physically
realistic with positive Brans-Dicke parameter, $\omega $, and describes either a
black hole or naked singularity:

\begin{equation}
ds^{2}=-dt^{2}(1-\frac{B}{r})^{m+1}-dr^{2}(1-\frac{B}{r})^{n-1}-r^{2}(1-%
\frac{B}{r})^{n}d\Omega ^{2},  \label{p}
\end{equation}

and the B-D scalar field, $\phi $, evolves as

\begin{equation}
\phi (r)=\phi _{0}(1-\frac{B}{r})^{-(m+n)/2}.  \label{q}
\end{equation}

\bigskip The BD coupling constant is related to $m$ and $n$ by

\begin{equation}
\omega =-2\left[ \frac{m^{2}+n^{2}+mn+m-n}{(m+n)^{2}}\right] .  \label{r}
\end{equation}

A black hole solution is allowed for $n<1$. The radial transformation $%
r\rightarrow \rho $ to isotropic coordinates defined by

\begin{equation}
r=\rho \left( 1+\frac{B}{r}\right) ^{2},  \label{s}
\end{equation}

reveals the more familiar form of the solution. Putting

\begin{eqnarray}
m &=&\frac{1}{\lambda }-1,  \label{t1} \\
n &=&1-\frac{C+1}{\lambda },  \label{t2}
\end{eqnarray}%
the scalar curvature invariant, $%
%TCIMACRO{\U{211d} }%
%BeginExpansion
\mathbb{R}
%EndExpansion
,$ is

\begin{equation}
%TCIMACRO{\U{211d} }%
%BeginExpansion
\mathbb{R}
%EndExpansion
=\frac{4\omega B^{2}C^{2}}{\lambda ^{2}\rho ^{4}(1+\frac{B}{\rho })^{8}}%
\left( \frac{1-B/\rho }{1+B/\rho }\right) ^{\frac{-2(2\lambda -C-1)}{\lambda
}}.  \label{u}
\end{equation}

When $n<-1$, i.e. for $2\lambda -C-1>0,$ $%
%TCIMACRO{\U{211d} }%
%BeginExpansion
\mathbb{R}
%EndExpansion
\rightarrow \infty $ as $\rho \rightarrow B$ and $\rho \rightarrow 0$ and
there is a naked singularity, But, when $(C+2-\lambda )/\lambda >0$ the
curvature invariants are non singular and $\rho =B$ is an event horizon.
However, this results in a violation of the weak energy condition because it
requires negative $\omega $, with
\begin{equation}
-2<\omega <-2(1+\frac{1}{\sqrt{3}}).  \label{v}
\end{equation}

In the naked singularity case the forces between point particles can become
arbitrarily large on approach to the singularity. In the black hole case the
only allowed black holes are the Schwarzschild black holes of GR \cite{swh}
and so the maximum force between them will still be $F_{GR}.$

\section{Dimensional features}

We have found earlier that there is a significant effect of spatial
dimension on the maximum force bounds. Only in three space dimensions is the
force bound independent of the masses of the gravitating objects. Moreover,
in $N$ space dimensions the fundamental classical 'Planck' quantity
depending in $G,$ and $c$ (but not $\hbar $) has dimensions of mass $\times $
(acceleration)$^{N}$, and is only a force when $N=3,$ see ref. \cite{JBsing},

\begin{equation}
MA^{N-2}=\frac{(F_{N})^{N-2}}{M^{N-3}}=\frac{(MA)^{N-2}}{M^{N-3}}=\left[
2(N-2)\right] ^{1-N}G_{{}}^{-1\ }\ c^{2(N-1)}.  \label{w}
\end{equation}

The maximum force between two touching $N$-dimensional black holes is

\begin{equation}
F_{N}=\ \frac{(N-2)8\pi GM^{2}}{(N-1)\Omega _{N-1}(2GM/c^{2})^{\frac{N-1}{N-2%
}}}\varpropto G^{\frac{-1}{N-2}}M^{\frac{N-3}{N-2}}c^{\frac{2(N-1)}{N-2\ }}.
\label{x}
\end{equation}

An interesting feature is the appearance of the mass of the attracting black
holes in higher-dimensional general relativity when (and only when) $N\neq3$. This may have some significance for the easier appearance of naked singularities in $N>3$ dimensions. However, we are interested to see if this feature persists in interesting generalisations of GR that retain second-order field equations, like versions of Lovelock's theory \cite{Love}. In the next section we will determine the existence of a maximum force in a preferred version of Lovelock's theory to see if it is mass-independent when $N>3.$

\section{Lovelock gravity}

In a $D$-dimensional spacetime, gravity can be described by an action
functional involving arbitrary scalar functions of the metric and curvature,
but not derivatives of curvature. In general, variation of such an arbitrary
Lagrangian would lead to an equation having fourth-order derivatives of the
metric. For them to be of second order, the gravitational lagrangian, $L$,
is constrained to be of the following Lovelock form, \cite{Love}:

\begin{equation}
L=\sum\lim_{n}c_{n}L_{n}^{{}}=c_{n}\frac{1}{2^{n}}\delta
_{c_{1}d_{1}c_{2}d_{2}....c_{n}d_{n}}^{a_{1}b_{1}a_{2}b_{2}...a_{n}b_{n}}R_{a_{1}b_{1}}^{c_{1}d_{1}}R_{a_{2}b_{2}}^{c_{2}d_{2}}....R_{a_{n}b_{n}}^{c_{n}d_{n}},
\label{love1}
\end{equation}%
where $\delta _{rs...}^{pq..}$ is the completely antisymmetric determinant
tensor. The case $n=1$ is the familiar Einstein-Hilbert lagrangian, while $%
n=2$ is the Gauss-Bonnet lagrangian, which is quadratic in curvature, and
reads

\begin{equation}
L_{2}\equiv L_{GB}=(1/2)(R_{abcd}R^{abcd}-4R_{ab}R^{ab}+R^{2}).
\label{love2}
\end{equation}

Lovelock's lagrangian is a sum over $n$, where each term is a homogeneous
polynomial in curvature and has an associated dimensionful coupling
constant, $c_{n}$. Moreover, the complete antisymmetry of the $\delta $
tensor demands $D\geq 2n$, or it would vanish identically. Even for $D=2n$
the lagrangian reduces to a total derivative. Lovelock's lagrangian, $L_{n},$
is therefore non-trivial only in dimension $D\geq 2n+1$.

Lovelock theory is the most natural and quintessential higher-dimensional
generalization of GR with the remarkable property that the field equations
continue to remain second order in the metric tensor despite the action
being a homogeneous polynomial in the Riemann tensor. GR is the linear order
Lovelock theory ($n=1$), whilst the Gauss-Bonnet term ($n=2$) is quadratic,
and then to any order, $n$, of the polynomial action. Each order comes with
a new arbitrary dimensionful coupling constant, $c_{n}$.

A particular minimal case of interest is that of the \textit{pure Lovelock}
which has only the $n^{th}$-order term in the lagrangian without a sum over
lower orders in the action and the equations of motion. It distinguishes
itself by the property that, as for GR in space dimension $N=2$, the
Lovelock analogue of the Riemann tensor \cite{cam-dad} is entirely given in
terms of the corresponding Ricci tensor in all critical even space
dimensions, $N=2n$. This property is in general termed 'kinematic'; for
example, GR is kinematic in $N=2\times 1=2$; Gauss-Bonnet would be kinematic
in $N=2\times 2=4$; that is, the pure Gauss-Bonnet Riemann\ tensor is
entirely given in terms of the corresponding Ricci, and this is so for all $%
N=2n,$\cite{cam-dad, dgj1}.

Pure Lovelock gravity is kinematic in all critical odd $D=2n+1$ dimensions
because the $n^{th}$ order Riemann tensor is entirely given in terms of the
corresponding Ricci tensor, hence it has no non-trivial vacuum solution.
Therefore, non-trivial vacuum solutions only exist in dimensions $D\geq 2n+2$%
. Finally, variation of the lagrangian with respect to the metric, for pure
Lovelock theories, leads to the following second-order equation,

\begin{equation}
-\frac{1}{2^{n+1}}\delta
_{c_{1}d_{1}c_{2}d_{2}....c_{n}d_{n}}^{a_{1}b_{1}a_{2}b_{2}...a_{n}b_{n}}R_{a_{1}b_{1}}^{c_{1}d_{1}}R_{a_{2}b_{2}}^{c_{2}d_{2}}....R_{a_{n}b_{n}}^{c_{n}d_{n}}=8%
\textcolor{red}{\pi} GT_{ab}.  \label{love3}
\end{equation}

Since no derivatives of curvature appear, this equation is of second order
in derivatives of the metric tensor. Although not directly evident, the
second derivatives also appear linearly and the equations are therefore
quasi-linear, thereby ensuring unique evolution.

Another property that singles out pure Lovelock is the existence of\emph{bound orbits} around a static object \cite{dgj2}. Note, that in GR,
bound orbits exist around a static object in Euclidean space only in three
space dimensions. In view of these remarkable features, it has been argued
that pure Lovelock is an attractive gravitational equation in higher
dimensions \cite{dad16}.

As with the Schwarzschild solution for GR, there exists an exact solution
for a pure Lovelock black hole \cite{dpp}, and it is given by %
Eq.(\ref{1a}), with (restoring the speed of light, $c$,
explicitly),

\begin{equation}
h(r)=1-2\Phi _{n}(r)/c^{2}  \label{y}
\end{equation}%
where the Newtonian potential term is

\begin{equation}
\Phi _{n}(r)=\frac{G_{n}M}{r^{\alpha }}, \quad \alpha =\frac{(D-2n-1)}{n}
\label{z}
\end{equation}

Here, $G_{n}$ is the gravitational constant for the $n^{th}$ Lovelock order,
$D=2n+1$ is the spacetime dimension ($D\neq 2n+1$ for a non-trivial vacuum
solution).

The analogue of the Newtonian inverse-square law now reads, for two equal
mass black holes of mass $M$:
\begin{equation}
F_{n}=\alpha G_{n}M^{2}/r^{\alpha +1}.  \label{z1}
\end{equation}

Then, $\Phi _{n}=c^{2}/2$ gives the black hole radius:

\begin{equation}
R_{g}=(2G_{n}M/c^{2})^{1/\alpha }.  \label{z2}
\end{equation}

Substituting this in the above force equation, we obtain,

\begin{equation}
F_{n}=\ \frac{\alpha G_{n}M^{2}}{(2G_{n}M/c^{2})^{(1+\alpha )/\alpha }}  \label{z3}
\end{equation}
which in the Planck units, analogue of Eq. (3), would read as
\begin{equation}
{F_{n}}_{pl}=\ G^{2/(1-N)} c^{(5+N)/(N-1)} h^{(3n-N)/(1-N)} .  \label{z33}
\end{equation}
Clearly, for $\alpha =1$, which implies $D=3n+1$ or $N=3n$, the maximum
force takes the same value, $c^{4}/4G_n$, as for GR in $N=3$ dimensions. Also the force in Planck units is free of Planck's constant in dimensions $N=3n$.  This
is because the pure Lovelock black hole potential goes as $1/r$ when $D=3n+1$
or $N=3n$. Note that this is an exact result, not an asymptotic one at large
$r,$ and is the same as the Newtonian potential for $N=3$ \cite{sum-dad}. In
particular, this also means that the maximum force value is mass and Planck's constant independent
in Lovelock gravity when $N=3n$. This feature only occurs in GR when $N=3$
and may have important indications for the problem of naked singularity
formation \cite{JBsing} because if the maximum force increases with $M$ then
it is possible for arbitrarily large forces to arise when sufficiently large mass
black holes interact. On the other hand the maximum force always has an upper bound for $N=3n$ in the pure Lovelock gravity which includes GR in the linear order $n=1$.%In $N=3$ then in GR, and in the pure Lovelock in general for $N=3n$ case we have analysed, this is not possible.
This is one
further remarkable property of pure Lovelock black holes and reveals the
effectiveness of the maximum force relation in analysing variants and
generalisations of GR. The recovery of the mass-independent maximum force
bound relies on the appearance of a $1/r$ gravitational potential in the
cases we have studied and is reminiscent of the conditions needed for the
Newton-Ivory spherical property and for \emph{closed bound} orbits in
Newtonian gravity \cite{JBnewt}.

\section{Conclusions}

We have searched for the existence of a maximum force in several gravity
theories that generalise GR in different ways. A maximum force $c^{4}/4G$ always exists for the pure Lovelock theory which includes GR in the linear order
%appears to exist generally in GR,
unlike in Newtonian gravity where forces
between points can become arbitrarily large as they approach. The GR and in general pure Lovelock maximum
force is independent of the attracting masses only in space dimensions three for GR and $3n$ for pure Lovlock theory. Our investigations find the following:

%a. In power-law lagrangians, the mass independence of the maximum force is achieved only for the case of the linear lagrangian ($\delta =0$), GR, and in all other cases, $\delta >0$, the maximum force is larger than in GR.

a. In Brans-Dicke theory, black holes are the same as in GR and the maximum
force is the same as in GR with mass independence in three space dimensions
only.

b. In Moffat's gravity theory there is a mass independent maximum force in
three space dimensions. The maximum force is smaller than in GR {and this
only happens in this case}.

c. In pure Lovelock gravity the maximum force is the same as in GR when the
space dimension is $3n$, where $n$ is the order of the one lagrangian
contributing to the Lovelock lagrangian. Therefore, the mass independent
maximum force bound exists in three dimensions for GR and Moffat's
generalisation of it, but in higher dimensions it only exists for pure
Lovelock with $N=3n$.

d. \emph{Modified gravity theories} generally introduced other arbitrary
dimensionful parameters into the lagrangian if there are additions to the
Einstein-Hilbert action. These can create new Planck-like quantities (see
ref.\cite{JBGib} for the impact of introducing a cosmological constant). Note that pure Lovelock theory avoids that as it has only single coupling constant.%We have avoided this in our power-law lagrangian example and in the pureLovelock lagrangian case, which has a single coupling constant.

It is rather remarkable that existence of mass independent maximum force singles out uniquely pure Lovelock from Lovelock theories. This is yet another distinguishing property in addition to gravity being kinematic in all critical odd $D=2n+1$ dimensions and existence of bound orbits around a static object. Though mass independence of maximum force occurs for Moffat's theory yet it however depends upon the Moffat's modification parameter $\alpha$. If we demand that the maximum force is entirely given in terms of velocity of light and gravitational constant, it is only the pure Lovelock theory  which is GR for $n=1$ that foots the bill and this is so in all $3n$ space dimensions.

 The recovery of the mass-independent maximum force bound relies on the
appearance of a $1/r$ gravitational potential in the cases we have studied
and is reminiscent of the conditions needed for the Newton-Ivory spherical
property and \emph{closed bound} orbits in Newtonian gravity \cite{JBnewt}.
We have not included a cosmological constant in our study, as was done in
ref. \cite{JBGib}, where a very similar mass-independent maximum force
bound, $c^{4}/9G$, was found in GR. \emph{Similar modifications of the bound
are to be expected in theories of gravity where new dimensionful parameters
are introduced into the gravity theory. }This is a direction for further
investigation in other gravity theories.

Acknowledgements. JDB is supported by the Science and Technology Funding
Council (STFC) of the UK and thanks Gary Gibbons for discussions.

\end{document}